# Experiments in Inferring Social Networks of Diffusion

Zoe Konrad and Daniel Campos

*Abstract*— Information diffusion is a fundamental process that takes place over networks. While it is rarely realistic to observe the individual transmissions of the information diffusion process, it is typically possible to observe when individuals first publish the information. We look specifically at previously published algorithm NETINF that probabilistically identifies the optimal network that best explains the observed infection times. We explore how the algorithm could perform on a range of intrinsically different social and information network topologies, from news blogs and websites to Twitter to Reddit.

## I. INTRODUCTION

Whether it is the spread of infectious diseases, the spread of news and opinions, or word of mouth effects in marketing, tracing a contagion as it is diffusing through the network is of interest to many researchers and marketers alike. Most typically, the network over which the propagation takes place is unknown and unobserved. We observe people getting sick but do not know who infected them; we observe people tweeting a new hashtag but do not know whose tweets they read to prompt it.

Our interest is to study paths of diffusion without complete information. Specifically the problem formulation we focus on is how to reconstruct the network of diffusion only observing the times when individual nodes get infected for each contagion. We build directly off of the work presented in the paper "Inferring Networks of Diffusion and Influence" by Gomez-Rodriquez, Leskovec, and Krause.

## II. NETINF

Gomez-Rodriquez, Leskovec, and Krause propose an algorithm called NETINF to solve this problem. NETINF tracks cascades of information diffusion among more than one million news media sites and blogs over a one year period. The algorithm efficiently reconstructs who-copies-from-whom network from these cascades. This makes it possible to see how different web sites copy from each other, and how a few central web sites have specific circles of influence.

The underlying basis for this algorithm is that by observing many different contagions spreading among the nodes, we can infer where edges are likely to be present in the underlying propagation network. If node *v* tends to get infected soon after node *u* for many different contagions, then we can expect an edge (*u*, *v*) to exist.

The NETINF algorithm goes contagion chain by contagion chain assembling probabilities for possible edges. The probability of edges is dictated by how quickly the contagion goes from one node to the next. Then, the algorithm selects the most probable selection of edges.

Each time the algorithm runs, it identifies the next most probable edge in the network. If the network runs 10 times, it identifies the 10 most probable edges. The more times you run the algorithm the more edges the algorithm will identify until it is not likely that any edges exist.

It is also worth noting that the space and runtime complexity for NETINF is impossible because it depends heavily on the structure of the network. NETINF exploits submodularity to make the inference problem tractable. This tractablility is a result of various optimizations and the fact that at each stage the algorithm only considers the most likely possibility.

The results of running the NETINF algorithm on real Web data are intuitive and clear. News media sites tend to diffuse news faster than blogs, and blogs keep discussing news for a longer time than media sites. The online news prorogation networks tend to have a core-periphery structure with a few blog and news sites diffusing information to the rest of the Web. These results are very exciting and show much promise for the real world applications of the NETINF algorithm.

Here, we choose to explore further what other applications might look like.

## III. EXPERIMENTAL EVALUATION

### A. Experiments with Generated Data

The first experiment was for us to familiarize ourselves with the NETINF algorithm and analyze results. We were interested in how some underlying properties of the network and parameters of the algorithm could affect the results.

Roughly following the experimental setup outlined in the original paper and using the SNAP library, we generated two random artificial Kronecker networks with 512 nodes and 1024 edges, thus very strongly connected. Giving each contagion a beta of 0.5 and an alpha assigned randomly ranging from 0.1 to 10, we generated 3 simulated sets of cascade chains of varying length where each node had an 85% chance of being infected within at least one cascade. The 3 sets had 256, 512 and 1024 cascade chains each.

Upon running the *NETINF* algorithm*,* we first observed that at lower iterations (say, less than 1,000), the edges of the inferred network estimated by the system were nearly identical. However, when the algorithm ran for more iterations than there were true edges, all of the simulations continued to find more edges than were actually present in the true network. So when the algorithm was run 2,000 times, 2,000 edges were found. This illustrates an important consideration of how the algorithm works for certain graph topologies. To the algorithm, if node u and v have an edge and node v and t have an edge, if we look long enough and our graph is connected enough, the algorithm will probably eventually conclude that nodes u and t have and edge, even if they do not.

This is not likely to occur in many real world networks, especially in the dataset we explore later in the paper, but it is important to keep in mind how the algorithm is applying its logic.

### B. Reddit Dataset

Our data set on Reddit is comprised of 132,308 image submissions from July 2009 until January 2013 with 16,736 unique images, each being submitted an average of 7.9 times. Each submission includes information such as subreddit, submitter id, votes and various timestamps. The data set was originally used to study the extent to which post title influence the success, measured by resubmissions and upvotes, of content in social media. We believed that this dataset would provide a very contrasted application for NETINF compared to the original news diffusion application.

One important distinction to make in Reddits infection model when compared to the models of data previously used with NETINF. Previous work treats epidemics as a SIR model where an individual is in one of three states (Susceptible, Infected, Recovered). We believe that data such as Reddit and other viral sharing follows more of a SI model (Susceptible, Infected) model. We do not believe that a user becomes infected, incubates and then infects beta of its peers. We believe that each user becomes infected with a virus (hashtag, image), and continues to infect its peers at a slower and slower rate. In other words, once a user is infected, it will continue to share that infection at varying rates. In our data, this matters because users often submit data to multiple Subreddits. These submissions do not commonly occur all at once. A common scenario may be user X submits an image and upon waiting some period of time they decide that another one of their communities would enjoy the image and they then post it to that community as well.

We expect the Reddit image submission to generally be much more sparse in the sense that for news, when there is a major event, you can count on quite a high percentage of people in the news media/blogosphere to post about it, whereas there are probably much fewer images with the same effect on Reddit. A big aspect here, though, is which data to use. For example, in the Memetracker application, the authors use a threshold where they only used memes that had cascade length of ten or more. This of course indicates that only cascades of that length add enough quality to justify including. This articulation of the tradeoff of quantity vs quality of data is of interest to us.

To get started with the experiments, we created a Python script to clean the data. First, we had to remove around 16,000 image submissions that lacked a username since knowing the node id is really the entire basis of the NETINF algorithm. Unfortunately, this results in a significant loss of the data and likely leaves many gaps in the network. Next, we identify cascades by common image id. The script then sorts the submissions in each cascade by time and convert the raw timestamp to the UNIX timestamp. Now our data in in proper form to be used as input for the algorithm: "node, time" for every contagion. We also allow for various tunable parameters in creating the dataset which allow us to further filter the data set by parameters such as minimum cascade length, vote count, etc.

So in the first steps of exploratory analysis, we wanted to see what threshold lengths would be reasonable for the Reddit data. We end up subsetted 19 different data sets by the minimum length of cascades present, from 2-20. Table 1 shows how thresholding based on cascade length impacts the quantity of data by amount of contagions (unique images) and transmissions (image postings). Our assumption is that for Reddit, longer cascades will be of higher value to helping the NETINF algorithm infer more edges of the network.

| Minimum Contagion | Average Length | Contagions | Transmissions |
|---|---|---|---|
| 2 | 6.9 | 16,727 | 114,313 |
| 3 | 7.3 | 16,541 | 111,611 |
| 4 | 9.0 | 15,190 | 99,260 |
| 5 | 10.5 | 8,405 | 88,588 |
| 6 | 12.1 | 6,563 | 79,378 |
| 7 | 13.7 | 5,214 | 71,824 |
| 8 | 15.2 | 4,255 | 64,571 |
| 9 | 16.7 | 3,512 | 58,627 |
| 10 | 18.0 | 2,994 | 53,965 |
| 11 | 19.4 | 2,545 | 49,475 |
| 12 | 21.0 | 2,156 | 45,196 |
| 13 | 22.4 | 1,856 | 41,596 |
| 14 | 23.9 | 1,606 | 38,346 |
| 15 | 25.5 | 1,381 | 35,196 |
| 16 | 26.9 | 1,214 | 32,691 |
| 17 | 28.4 | 1,074 | 30,451 |
| 18 | 29.8 | 954 | 28,441 |
| 19 | 31.0 | 866 | 26,827 |
| 20 | 32.6 | 762 | 24,851 |

### C. Environment Setup

Initially, our experiments began on small laptops but we soon learned that was not possible nor scalable. We did not have enough RAM causing most programs crashed on larger amounts of iterations. We also had few and slow cores causing inhibiting our ability to run multiple experiments simultaneously. Knowing this, we decided to move to the cloud, specifically Microsoft's Azure platform, which we happened to have some free credit for.

Our plan was to run our 19 data sets through NETINF using the following iterations: 100, 250, 500, 1000, 2000, 5000, 10000, 50000. At lower iterations we expected that we would have many sparse clusters. We thought that as we increased the iterations, we would achieve more and more connected graphs and get more of a clue to how users interacted and perhaps could group users together via Subreddit or 'Viral' users.

In Azure, we experimented with various builds of OS ranging from Windows Server 2012 to various versions of Ubuntu and finally setting on the most recent Ubuntu 14.10 hoping that is thread usage and RAM usage would be the best. We knew we needed a lot of ram so we used Azure's D14 servers, which have 16 high performance cores and 112 GB of ram. From there we began to run our experiments using scripts that ran each test one at a time. Running on a single core, each ranges from around 15 minutes (20 transmissions per cascade or more) to 70 minutes (no minimum). This initial setup would take approximately 150 hours to run and on average CPU usage was around 12%.

With the slow experiments running, we moved on to optimizations by creating more server instances and modifying our testing scripts. At first, we tried running segmented versions of the script simultaneously. In other words, instead of having a script that ran all the iterations we spit we used 8 scripts, each which each ran a specific iteration size on the dataset. This improved both our time (around 25 hours) and increased out CPU utilization (35%). This was faster but the biggest issue was that not all iterations took the same amount of time (100 iterations is much faster than 50,000) and left idle CPU and occasion caused crashed due to RAM issues (Running all the longer cascades at once).

The next changes we made were to modify the scripts to make batches of documents. In other words one core would get documents 1,3 another would get 2,14. This solution increased the compute time (around 30 hours) and decreased CPU usage. This was because not all sets were balanced by how long they would take and if two bigger ones got paired together, that core would perform even worse.

Our final script that we tried generated all the commands and then dispatched them all at random at cores became free. This system was by far the most effective, we essentially saw a total run time of about 8 hours and our CPU usage was close to 99% the entire time. Further improvements may be to merge reading of files and improve how information is stored in memory to decrease lookups and crashes.

D. *Running NETINF on Reddit Dataset*

Upon finishing our tests we visualized the graphs and tried to see what overall information we could infer. We were surprised to see most of edges in the data were unidirectional and form chains of 3 or 4 edges. The next thing we noticed was the prevalence of a few 'Hub' posters. These posters were usually characterized by a high amount of inward and outward edges. These users served as aggregators, consuming a lot of content and then dispersing these images to other Subreddits. What is not possible to know about these users is their success rate. Since Reddit users value the websites currency, karma, it is possible that these users become infected by thousands of images but very few of them succeed.

Analyzing the large scale results produced an answer opposite of our hypothesis. Instead of producing more edges with our higher quality cascade chains, they produced less and had less connected graphs. The results that produced the most information and the most useful insights were those with the highest amount of transmissions. We also found that our results did not behave at all like our generated data. Regardless of how many iterations we ran the algorithm for, each document produced an upper bound for how many edges could be inferred. This is completely contrasting to our generated data that would create more edges than really existed. This is good to know for future work as we do not have to worry about this issue when dealing with network identification in large, diverse social networks.

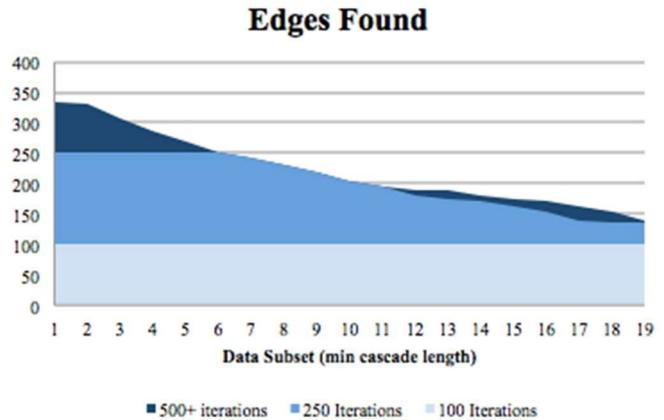

*Figure 1: number of edges in the inferred network by data subset and iterations*

Next, looking at the following figures, it is interesting to notice the edges that were found by our datasets with more cascades are more 'vital' to the connectivity as a whole. Without these edges inferred from this extra data, the graphs appear as mostly a jumbled mess. We believe that if we had more contagions, even if they were short, we would be able to find even more information about the network. We believe that these short cascades may even be more useful to the system as there are less possible people who affected the infection of shorter nodes.

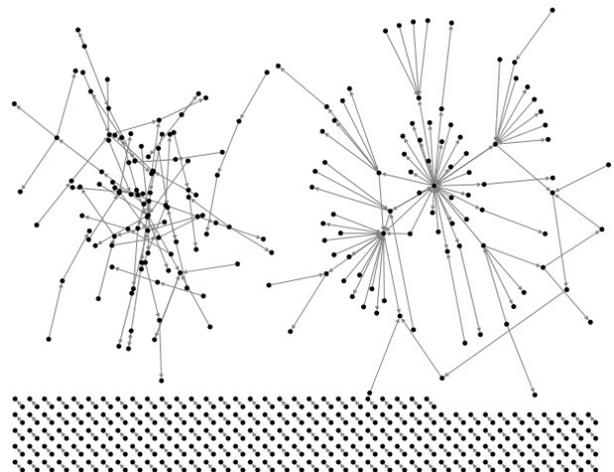

*Figure 2: A visualization of the inferred Reddit network with no minimum cascade length*

In Figure 2 there are clearly clusters. We believe that if we had access to more data, we would find many such 'hub' users that would be connected through weak links. It is also important to note the frailty of these graphs. If we removed a few cascades that included that central user, out graph would appear unconnected, like we see in Figure 4.

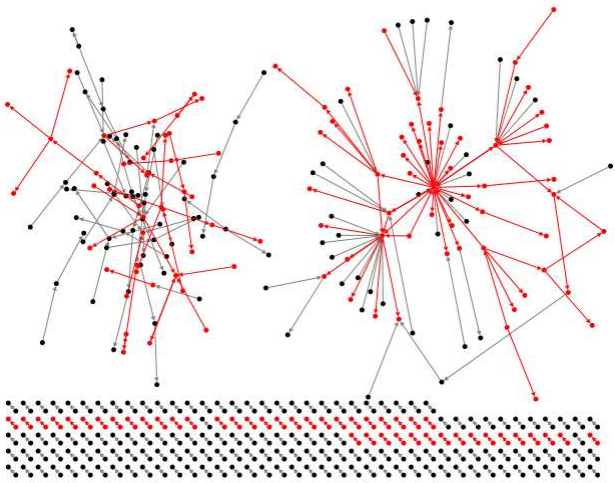

*Figure 3: A visualization of the inferred Reddit network after introducing the minimum cascade length of 19. Red edges are the edges that are not discovered compared to using no minimum.*

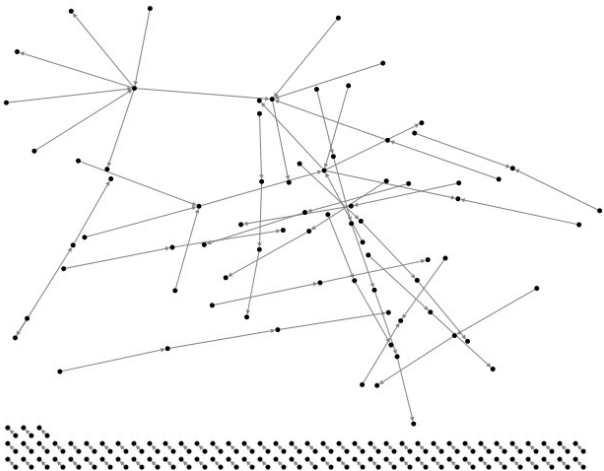

*Figure 4: A visualization of the inferred Reddit network with minimum cascade length of 19.*

We believe there are a few reasons why we were only able to draw such few inferences about the Reddit community as a whole. The first reason is lack of data. While the data set is relatively large, it only represents a small sample of the data present in Reddit. This data was originally used to provide a sampling of data and while this produces an accurate display of the breadth of topics, it lacks depth. To have more thorough coverage of Reddit we would likely need data 1 or 2 magnitudes of scale larger. When we compare the size of this data set to ones previously used with NETINF that are in related sphere (Memetracker) Reddit is about 10% of the size.

A second reason would be related to the actual network topology of Reddit. Reddit is made up of many small networks that are not strongly connected. In other words, there is little overlap between the communities for kitten lovers and death metal bands. Moreover, even if there is overlap, these users who are active in separate networks, they are unlikely to share items across networks since the posts tend to be very focused.

While this survey of Reddit using NETINF produced little information about the underlying network topology it identified a few factors about cascades and target networks that we plan to use in our continued work.

## IV. FUTURE WORK

Our attempts to extend the practical applications of the NETINF algorithm were extremely limited by both time and lack of availability of quality datasets.

The major accomplishment of our work here is the solid implementation foundation we established. We have the entire process up and running from cleaning the data to running the NETINF algorithm efficiently in the cloud.

The first area improvement is working on making the algorithm faster and setting up a system to run this algorithm quickly. Since each set of probabilities is calculated for each cascade independently, the problem is perfect for some kind of concurrent programming solution such as actor based programming. We are evaluating how to set up a system that would be able to run in parallel, hopefully allowing the system to take full advantage of cheap CPU and scalability.

A second major extension would be applying this data set to other viral data such as Twitter, which would provide a unique opportunity to compare real network topology results in two ways. We could first model out network data using edges we know to exist, like retweet chains. Then, we would run our system using various contagions such as hashtags or memes to infer the network. We would be able to compare the networks in a unique way similar to using blog hyperlinks and Memetracker to infer news diffusion networks.

We have an optimistic outlook for the future work building off of this foundation. A top area of interest for us is conducting more robust experiments in tradeoffs between performance time and accuracy for certain parameters of the algorithm. The goal would be to make this code available along with comprehensive recommendations of parameter settings based on characteristics of a dataset and performance needs of the user.